\documentclass[aps,prfluids,preprint,superscriptaddress,nofootinbib]{revtex4-2}

\usepackage{amsmath}
\usepackage{amssymb}
\usepackage{graphicx}
\usepackage{hyperref}
\usepackage{subcaption}
\usepackage{overpic}
\begin{document}

\title{A conservative low-order model for Boussinesq baroclinic fronts}

\author{Nadav Yovel}
\affiliation{School of Physics and Astronomy, Tel Aviv University, Tel Aviv 69978, Israel}

\author{Eyal Heifetz}
\email{eyalh@tauex.tau.ac.il}
\affiliation{Department of Geophysics, Tel Aviv University, Tel Aviv 69978, Israel}

\date{\today}

\begin{abstract}
The internal dynamics of baroclinic fronts are governed by a fundamental interplay: turbulent eddies systematically act to disrupt thermal wind balance, with baroclinic eddies flattening isopycnals and barotropic momentum fluxes intensifying the primary jet, while the ageostrophic overturning circulation acts to restore it. In quasi-balanced models, this restorative adjustment is assumed to be instantaneous, effectively locking the flow onto a balanced manifold between the eddies and the overturning cell. To conceptually track this physical mechanism when the adjustment takes a finite time, we construct a low-order model that spans from $\mathcal{O}(1)$ Rossby numbers down to the quasi-balanced limit. Formulated consistently from the continuous Boussinesq equations under a $Ro^2 Ri \sim 1$ scaling, which inherently constrains the horizontal length scale to the Rossby deformation radius, the derivation yields a closed, nonlinear five-dimensional ordinary differential equation system. The degrees of freedom consist of the domain-averaged along-front vertical shear, the cross-frontal overturning vorticity, the horizontal and vertical buoyancy gradients, and the total eddy energy. We identify two constants of motion that constrain the evolution of the mean flow: the total energy (comprising the kinetic energy of the along- and cross-frontal flows, mean potential energy, and eddy energy) and the magnitude of the domain-averaged cross-frontal density gradient. Notably, while the system is energetically conservative, the parameterized turbulent closure renders the dynamics strictly non-Hamiltonian. Bounded by these invariants, the adiabatic adjustment of the front physically reduces to a continuous rotation of the density gradient's slope. By explicitly resolving the inertial lag of the secondary circulation, this framework allows us to isolate the individual mechanisms governing frontal adjustment and track their continuous dynamic interplay.
\end{abstract}

\maketitle

\section{Introduction}
\label{sec:intro}

Baroclinic fronts are ubiquitous features across a vast range of spatial scales in both the atmosphere and the ocean, ranging from planetary-scale atmospheric jet streams \cite{Vallis2017, Hoskins1978} to mesoscale and submesoscale filaments \cite{McWilliams2016, Thomas2008}. They are fundamentally characterized by strong cross-frontal density gradients. To maintain equilibrium, the primary along-front flow resides in a state of thermal wind balance, where the vertical shear of the jet is dynamically supported by the lateral buoyancy gradient (Figs.~\ref{fig:Slope_Struc}, \ref{fig:dynamical_system}a).

The internal dynamics of these fronts, however, are governed by a continuous interplay of competing mechanisms. On one hand, turbulent eddies systematically act to disrupt the thermal wind balance. Baroclinic eddies extract available potential energy, effectively flattening the isopycnals and relaxing the cross-frontal gradient \cite{FoxKemper2008}, while barotropic eddy momentum fluxes often converge to accelerate and sharpen the mean jet \cite{James1987} (Fig.~\ref{fig:dynamical_system}b). On the other hand, this turbulent disruption necessitates the generation of an ageostrophic cross-frontal overturning circulation (Fig.~\ref{fig:dynamical_system}c), which acts to restore the thermal wind balance and heal the kinematic rupture (Fig.~\ref{fig:dynamical_system}d).

In quasi-balanced theories of large-scale geophysical fluid dynamics, such as quasi-geostrophy (QG), the analysis of this adjustment relies on a strict scale separation indicated by a small Rossby number ($Ro \ll 1$) \cite{Vallis2017}. A direct mathematical consequence of the $Ro \to 0$ limit is the filtering of inertio-gravity waves and inertial oscillations. The fluid is thereby assumed to adjust instantaneously, and the flow is effectively locked onto a balanced manifold between the eddies and the overturning cell. The diagnosis of the secondary circulation is thus reduced to the inversion of a spatial elliptic operator. This is elegantly captured by the classical Sawyer-Eliassen equations and the Q-vector formulation for frontal zones \cite{Sawyer1956, Eliassen1962, Hoskins1978}, as well as by the Kuo equation for the planetary-scale Ferrel cell \cite{Kuo1956}.

However, when the horizontal length scale remains the deformation radius but the Rossby and Richardson numbers are of order one ($Ro \sim 1, Ri \sim 1$), this elliptic assumption breaks down \cite{Thomas2008}. The adjustment takes a finite time, and the secondary circulation exhibits an inertial lag. Intermediate theoretical frameworks offer partial bridges to this regime. For instance, semi-geostrophic theory successfully extends balanced dynamics to capture finite-amplitude frontal steepening, yet it fundamentally retains the assumption of instantaneous cross-frontal adjustment \cite{Hoskins1975}. Similarly, while nonlinear multi-layer shallow-water models accurately resolve inertial lag and finite-time frontal evolution, their formulation relies on discrete density interfaces, which obscures the continuous thermodynamic rotation of the bulk stratification. Consequently, while high-resolution continuous Primitive Equation (PE) simulations capture these finite-amplitude dynamics accurately \cite{OSullivan1995, Plougonven2007}, their structural complexity makes it difficult to conceptually track the individual terms governing the adjustment mechanism and their dynamic interplay.

To bridge this gap, the objective of this paper is to construct a low-order, nonlinear set of ordinary differential equations (ODEs) that spans from $\mathcal{O}(1)$ Rossby numbers down to the quasi-balanced limit. Extending the recent Hamiltonian formalisms developed for horizontally convective basins \cite{Maas2026}, we systematically reduce the continuous Boussinesq momentum and thermodynamic equations into a fully conservative system with five degrees of freedom. We demonstrate that this reduction does not rely on arbitrary dynamic filtering; rather,  nonlinear interactions, such as the kinematic self-deformation of the secondary circulation, vanish exactly due to mass continuity and the spatial symmetries of the frontal domain. Bounded by exact constants of motion, we show that the complex adiabatic adjustment of the front can be mapped to a continuous physical rotation of the density gradient's slope.

The paper is organized as follows. In Section \ref{sec:boussinesq}, we formulate the continuous Boussinesq equations and define the bulk geometry and domain-averaged variables. In Section \ref{sec:eddy_energy}, we apply Reynolds decomposition and establish the energetic truncation and closure to isolate the dominant instabilities. In Section \ref{sec:derivation}, we perform the spatial projection and non-dimensionalization, deriving the fully closed 5D ordinary differential equations for the frontal system. In Section \ref{sec:properties_interpretation}, we evaluate the dynamical properties and constants of motion of the model, and detail the physical interpretation of the inherent, self-regulating feedbacks. In Section \ref{sec:discussion}, we trace a canonical physical scenario to illustrate the continuous chronological adjustment of the front. Finally, conclusions are drawn in Section \ref{sec:conclusion}, where we outline the broader implications of the framework, marking the path for future investigations.

\section{Boussinesq Framework and Bulk Geometry}
\label{sec:boussinesq}

\subsection{The Governing PDEs and Density Decomposition}

We begin our derivation with the continuous, inviscid, and adiabatic Boussinesq equations on an $f$-plane. For an incompressible fluid ($\nabla \cdot \vec{u} = 0$), the three-dimensional momentum and thermodynamic equations read:
\begin{subequations}\label{eq:boussinesq}
\begin{align}
    \frac{D\vec{u}}{Dt} + f \hat{k} \times \vec{u} &= -\nabla \phi + b \hat{k} \, , \label{eq:boussinesq_mom} \\
    \frac{Db}{Dt} + N^2 w &= 0 \, , \label{eq:boussinesq_buoy}
\end{align}
\end{subequations}
where $\frac{D}{Dt} = \frac{\partial}{\partial t} + \vec{u} \cdot \nabla$ is the material derivative, and $\vec{u} = (u,v,w)$ is the total velocity vector. 

To formally define the dynamic buoyancy $b$ and the background stratification $N$, we decompose the total density field as $\rho = \rho_m + \rho_0(z) + \overline{\rho}(y,z,t) + \rho'(x,y,z,t)$, where $\rho_m$ is the constant reference density, $\rho_0(z)$ is the static background density profile, $\overline{\rho}$ is the dynamic zonal-mean anomaly, and $\rho'$ represents the eddy fluctuations ($\overline{\rho'} = 0$). 

To map these components to the Boussinesq framework, we define the background static buoyancy $b_0(z) \equiv -g\rho_0/\rho_m$ and the mean dynamic buoyancy anomaly $\bar{b}(y,z,t) \equiv -g\overline{\rho}/\rho_m$. The total dynamic buoyancy driving the momentum equation is thus $b = \bar{b} + b'$, where $b'$ is the eddy buoyancy anomaly. The static background profile defines the constant, stable background stratification, $N^2 \equiv db_0/dz = -(g/\rho_m)(d\rho_0/dz)$, represented by the square of the Brunt-V\"ais\"al\"a frequency $N$. 

Finally, the total pressure is partitioned as $p_{tot} = p_0(z) + p$. The background pressure $p_0(z)$ completely absorbs the static hydrostatic balance, $\frac{1}{\rho_m} \frac{d p_0}{dz} = -\left(1 + \frac{\rho_0(z)}{\rho_m}\right)g$, leaving $\phi \equiv p/\rho_m$ to represent the kinematic pressure deviation from this reference.

\subsection{Domain Averaging and Bulk Variables}
\label{subsec:Bulk}
We consider the front to be aligned along the zonal direction ($x$) and assume it to be zonally uniform ($\partial/\partial x = 0$). This two-dimensional meridional-vertical $(y,z)$ framework captures cross-frontal overturning dynamics, making the model applicable to generic elongated fronts as well as planetary-scale zonal circulations. Locating the front at $y=0$, we consider a cross-frontal circulation cell defined by $|y| \le L/2$ and $0 \le z \le H$. The domain average over this cell is defined as:
\begin{equation}\label{eq:domain_average}
    \langle ... \rangle \equiv \frac{1}{HL} \int_{-L/2}^{L/2} \int_{0}^{H} (...) \, dy \, dz \, .
\end{equation}

The cross-frontal overturning circulation in the $(y,z)$ plane generates a domain-averaged zonal vorticity component: $l \equiv \langle \frac{\partial \bar{w}}{\partial y} - \frac{\partial \bar{v}}{\partial z} \rangle$. Because of the zonal uniformity ($\partial \bar{w}/\partial x = 0$), the circulation in the $(x,z)$ plane is unclosed. Consequently, the domain-averaged meridional vorticity results solely from the vertical shear of the along-front velocity: $m \equiv \langle \frac{\partial \bar{u}}{\partial z} \rangle$.

For simplicity, we assume the mean dynamic density anomaly, $\overline{\rho}(y,z,t)$, has a linear spatial structure across the frontal zone. This geometrically dictates that its spatial gradients are uniform and depend exclusively on time:
\begin{equation}\label{eq:density_structure}
    \overline{\rho}(y,z,t) = \frac{\partial \overline{\rho}}{\partial y}(t)\, y + \frac{\partial \overline{\rho}}{\partial z}(t)\, z \, .
\end{equation}
Applying the Boussinesq mapping ($\bar{b} \equiv -g\overline{\rho}/\rho_m$), this directly yields a linear spatial structure for the mean dynamic buoyancy:
\begin{equation}\label{eq:buoyancy_structure}
    \bar{b}(y,z,t) = -Y(t)\, y + \bar{b}_z(t)\, z \, ,
\end{equation}
where $Y(t) \equiv -\partial \bar{b}/\partial y = (g/\rho_m)\partial \overline{\rho}/\partial y$ represents the bulk horizontal buoyancy gradient, and $\bar{b}_z(t) \equiv \partial \bar{b}/\partial z = -(g/\rho_m)\partial \overline{\rho}/\partial z$ is the dynamic vertical buoyancy anomaly resulting from the front.
We define the total macroscopic vertical stratification $Z(t)$ as the sum of the background static stability ($N^2$) and this dynamic anomaly. Because $N^2 \equiv db_0/dz$, this formally ensures that $Z$ represents the vertical gradient of the total mean buoyancy ($b_{\text{tot}} \equiv b_0 + \bar{b}$):
\begin{equation}\label{eq:stratification_structure}
    Z(t) \equiv N^2 + \bar{b}_z(t) = \frac{\partial b_{\text{tot}}}{\partial z} \, .
\end{equation}
As $N^2$ is constant, the evolution of these macroscopic variables directly reflects the spatial gradients of the local dynamic buoyancy tendency: $(\dot{Y}, \dot{Z}) = \left(-\frac{\partial}{\partial y}, \frac{\partial}{\partial z}\right)\dot{\bar{b}}$.

The geometric center of the meridional cell is $(y,z)_{\text{g.c.}} = (0, H/2)$, whereas the center of mass is given by $(y,z)_{\text{c.m.}} = (\langle \rho y \rangle, \langle \rho z \rangle)/\rho_m = \left(\frac{L^2}{12 g}Y, \frac{H}{2} - \frac{H^2}{3 g}Z\right)$. Thus, when $Y > 0$, denser fluid lies to the north of the front. We assume the flow remains stably stratified (dominated by $N^2$, so that $Z > 0$ at all times); consequently, large values of $Z$ lower the vertical center of mass well below the geometric center. The cross-frontal isopycnal slope of the total density field is exactly $S(t) = Y/Z$, and the magnitude of the macroscopic cross-frontal density gradient is $|\nabla \rho_{\text{total}}| = \frac{\rho_m}{g}\sqrt{Y^2 + Z^2}$ (Fig.~\ref{fig:Slope_Struc}).

Finally, for the primary along-front jet $\bar{u}(y,z)$, we assume the flow is centered within the domain at $y=0$. Consequently, the jet velocity profile is symmetric in the meridional direction, rendering its horizontal shear $\bar{u}_y$ anti-symmetric. While the exact domain average of this lateral shear vanishes ($\langle \bar{u}_y \rangle = 0$), its characteristic magnitude across the baroclinic zone scales proportionally to the primary vertical shear $m$ via the geometric aspect ratio of the front, scaling dynamically as $(H/L)m$.

\section{Energetic Truncation and Eddy Energy Closure}
\label{sec:eddy_energy}

\subsection{Energy Budget}

To establish the energy budget governing the growth of perturbations, we first examine the conservation of total mechanical energy. Taking the dot product of $\vec{u}$ with Eq.~\eqref{eq:boussinesq_mom}, noting that the Coriolis force performs no work ($\vec{u} \cdot (f \hat{k} \times \vec{u}) = 0$), and using the divergence-free condition to rewrite the pressure gradient term ($\vec{u} \cdot \nabla \phi = \nabla \cdot (\vec{u}\phi)$), we obtain the kinetic energy tendency:
\begin{equation}\label{eq:kin_energy}
    \frac{D}{Dt}\left(\frac{\vec{u}^2}{2}\right) = -\nabla \cdot (\vec{u}\phi) + wb \, .
\end{equation}

To close the energy loop, we isolate the vertical velocity from the buoyancy equation \eqref{eq:boussinesq_buoy}, $w = -N^{-2} Db/Dt$, and substitute it into the buoyancy work term $wb$. This allows us to express the available potential energy density directly as $P \equiv b^2 / (2N^2)$. Defining the kinetic energy density as $K \equiv \vec{u}^2 / 2$, and substituting these into Eq.~\eqref{eq:kin_energy}, yields the standard conservation of total mechanical energy for a Boussinesq fluid parcel \cite{Vallis2017}:
\begin{equation}\label{eq:mech_energy}
    \frac{D}{Dt}\left(\frac{\vec{u}^2}{2} + \frac{b^2}{2N^2}\right) = -\nabla \cdot (\vec{u}\phi) \, .
\end{equation}

Applying a full three-dimensional volume average, achieved by first taking the zonal mean (denoted by the overbar $\overline{(\cdot)}$) followed by the 2D cross-frontal domain average ($\langle\cdot\rangle$) of eq.~\eqref{eq:domain_average}, the divergence theorem dictates that the boundary pressure fluxes vanish. This confirms the conservation of total volume-averaged energy: $\frac{\partial}{\partial t} \langle \overline{K + P} \rangle = 0$. Given the shallow aspect ratio of the fluid domain, $(H/L)^2 \ll 1$, the vertical kinetic energy is geometrically suppressed and thus assumed negligible. We therefore restrict our definition of the kinetic energy density strictly to the horizontal velocity components: $K \simeq (u^2 + v^2)/2$.

\subsection{Reynolds Decomposition and Energy Generation}

To separate the dynamics of the along-front jet and the overturning circulation from the eddy field, we apply Reynolds decomposition. Variables are partitioned into a zonally uniform mean ($\partial/\partial x = 0$) and three-dimensional eddy fluctuations: $\vec{u} = \bar{\vec{u}}(y,z,t) + \vec{u}'(x,y,z,t)$ and $b = \bar{b}(y,z,t) + b'(x,y,z,t)$. 

Because the domain average of the linear perturbation terms vanishes ($\langle \vec{u}' \rangle = 0$ and $\langle b' \rangle = 0$), the cross-terms between the mean and eddy fields (such as $\langle \bar{\vec{u}} \cdot \vec{u}' \rangle$ and $\langle \bar{b}b' \rangle$) in the squared quantities integrate to zero. The total energy tendency naturally decouples into mean and eddy components:
\begin{equation}\label{eq:energy_sep}
  \dot{e} \equiv \left\langle \frac{\partial}{\partial t}\left( \frac{u'^2 + v'^2}{2} + \frac{b'^2}{2N^2} \right) \right\rangle = -\left\langle \frac{\partial}{\partial t}\left( \frac{\bar{u}^2 + \bar{v}^2}{2} + \frac{\bar{b}^2}{2N^2} \right) \right\rangle = -\left\langle \bar{u}\dot{\bar{u}} + \bar{v}\dot{\bar{v}} + \frac{\bar{b}\dot{\bar{b}}}{N^2} \right\rangle  \, .
\end{equation}

To evaluate this, we explicitly write the zonally averaged ($\partial/\partial x = 0$) momentum and buoyancy equations for the mean basic state. Denoting the cross-frontal circulation vector as $\bar{\vec{v}} = (0, \bar{v}, \bar{w})$ and the 2D cross-frontal gradient operator as $\bar{\nabla} = (0, \partial/\partial y, \partial/\partial z)$, the tendencies are:
\begin{subequations}\label{eq:mean_tendencies}
\begin{align}
    \frac{\partial \bar{u}}{\partial t} &= -\bar{\nabla} \cdot (\bar{\vec{v}}\bar{u}) + f\bar{v} - \bar{\nabla} \cdot (\overline{\vec{v}' u'}) \, , \label{eq:mean_u} \\
    \frac{\partial \bar{v}}{\partial t} &= -\bar{\nabla} \cdot (\bar{\vec{v}}\bar{v}) - f\bar{u} - \frac{\partial \bar{\phi}}{\partial y} - \bar{\nabla} \cdot (\overline{\vec{v}' v'}) \, , \label{eq:mean_v} \\
    \frac{\partial \bar{b}}{\partial t} &= -\bar{\nabla} \cdot (\bar{\vec{v}}\bar{b}) - N^2 \bar{w} - \bar{\nabla} \cdot (\overline{\vec{v}' b'}) \, . \label{eq:mean_b}
\end{align}
\end{subequations}
Here, we have expressed the mean advection terms in conservative flux form, utilizing the incompressibility condition: $\nabla \cdot \vec{u} = 0 \implies \bar{\nabla} \cdot \bar{\vec{v}} = 0$.

Substituting Eqs.~\eqref{eq:mean_tendencies} into Eq.~\eqref{eq:energy_sep}, the products of the mean variables and their advective fluxes can be rewritten as the divergence of their quadratic fluxes (e.g., $\bar{u} \bar{\nabla} \cdot (\bar{\vec{v}}\bar{u}) = \bar{\nabla} \cdot (\bar{\vec{v}} \bar{u}^2 / 2)$). Integrating over the closed domain, these conservative quadratic fluxes evaluate to zero via the divergence theorem, as there is no normal flow across the boundaries. 

Furthermore, recognizing that the mean basic state is in hydrostatic balance ($\bar{b} = \partial \bar{\phi} / \partial z$), the Coriolis and ageostrophic cross-terms exactly cancel ($\langle f\bar{u}\bar{v} - f\bar{u}\bar{v} \rangle = 0$ and $\langle -\bar{v}\bar{\phi}_y - \bar{w}\bar{b} \rangle = -\langle \bar{\vec{v}} \cdot \bar{\nabla} \bar{\phi} \rangle = -\langle \bar{\nabla} \cdot  (\bar{\vec{v}} \, \bar{\phi}) \rangle =  0$). This leaves the evolution of eddy energy governed by the divergence of the eddy momentum and buoyancy fluxes acting on the basic state. Integrating these remaining terms by parts yields the expression for the eddy energy generation:
\begin{equation}\label{eq:e_dot_full}
    \dot{e} = -\langle \overline{u'v'}\bar{u}_y + \overline{u'w'}\bar{u}_z + \overline{v'v'}\bar{v}_y + \overline{v'w'}\bar{v}_z \rangle - \langle \overline{v'b'}\frac{\bar{b}_y}{N^2} + \overline{w'b'}\frac{\bar{b}_z}{N^2} \rangle \, .
\end{equation}

While this expression is exact, we apply a specific energetic truncation to isolate the dominant frontal-scale instabilities of the front. In the $\mathcal{O}(1)$ Rossby number regime characteristic of submesoscale dynamics, vertical Reynolds stresses ($\overline{u'w'}$) can be of the same order as the meridional stresses. However, our primary focus is on lateral instability pathways rather than small-scale, Kelvin-Helmholtz-like vertical shear instabilities. Furthermore, as the Rossby number decreases toward the quasi-balanced limit, this vertical conversion pathway naturally ceases to be the main energy conduit. Thus, we explicitly filter out kinetic energy conversion via the mean vertical shear ($\overline{u'w'}\bar{u}_z$). 

Second, while the secondary circulation meridional flow ($\bar{v}$) can be highly energetic at $Ro \sim \mathcal{O}(1)$, there is no established physical mechanism to spontaneously activate its horizontal eddy momentum convergence. Unlike the primary along-front jet, which is systematically intensified by the upscale transfer of kinetic energy (e.g., via submesoscale vortex merging and nonlinear scale interactions), there is no equivalent upscale cascade projecting strictly onto the cross-frontal ageostrophic direction. We therefore treat the overturning cell fundamentally as an ageostrophic restoring response, neglecting the eddy momentum work acting directly upon it ($\overline{v'v'}\bar{v}_y$ and $\overline{v'w'}\bar{v}_z$).

Finally, regarding the available potential energy, the energetic reservoir of the front is governed by the lateral density gradient. The subdominance of the vertical eddy buoyancy flux ($\overline{w'b'}\bar{b}_z / N^2$) can be demonstrated explicitly. For baroclinic instability to successfully extract APE, the eddy trajectories must slope within the baroclinic wedge of instability (i.e., bounded between the horizontal plane and the isopycnals). This kinematically constrains the eddy velocity ratio $w'/v'$ to scale with the isopycnal slope, $S = -\bar{b}_y / (\bar{b}_z + N^2)$. Under the standard assumption that the background stratification dominates the dynamic vertical anomaly ($\bar{b}_z \ll N^2$), this slope approximates to $-\bar{b}_y / N^2$. Evaluating the ratio of the vertical to the lateral buoyancy work thus yields $(w'/v')(\bar{b}_z / \bar{b}_y) \sim (-\bar{b}_y/N^2)(\bar{b}_z/\bar{b}_y) = -\bar{b}_z/N^2$. This resulting small parameter mathematically verifies that the vertical term is strictly subdominant to the primary lateral extraction ($\overline{v'b'}\bar{b}_y / N^2$).

Applying this dynamic truncation reveals the unifying physical feature where both the kinetic and available potential energy conversions are governed by the meridional eddy fluxes:
\begin{equation}\label{eq:e_dot_simplified}
    \dot{e} = -\langle \overline{u'v'}\bar{u}_y \rangle - \langle \overline{v'b'}\frac{\bar{b}_y}{N^2} \rangle \, .
\end{equation}

\subsection{Energetic Closure}

To close Eq.~\eqref{eq:e_dot_simplified}, we first establish the physical direction of these two distinct energy transfer pathways based on established frontal dynamics.

The barotropic conversion term, $-\langle \overline{u'v'}\bar{u}_y \rangle$, dictates the lateral exchange of kinetic energy. In typical baroclinic fronts, this term acts as an energetic sink for the turbulent field, as the eddy momentum covariance structurally phase-locks to intensify the primary along-front jet. At the submesoscale ($Ro \sim \mathcal{O}(1)$), such as during mixed layer instability (MLI), this intensification is driven by an upscale kinetic energy cascade via vortex merging \cite{Boccaletti2007}. As the spatial scales expand toward the classical quasi-geostrophic limit ($Ro \to 0$), this upscale transfer is maintained by the far-field radiation of Rossby waves, which induces a meridional convergence of eddy momentum \cite{Hoskins1983}. Consequently, we assume this global kinetic energy exchange is strictly negative.

Conversely, the baroclinic conversion term, $-\langle \overline{v'b'}\frac{\bar{b}_y}{N^2} \rangle$, serves as the primary energy source for the eddies. Driven by baroclinic instability, the turbulent fluxes systematically act to release mean available potential energy by flattening the tilted isopycnals, guaranteeing a strictly positive energy conversion.

Before parameterizing the turbulent fluxes, we project the mean spatial gradients onto the domain-averaged variables. Using the bulk geometry established in Section \ref{sec:boussinesq}, the lateral shear relates to the domain-averaged vertical shear, $m$. Because the primary jet is centrally localized within the frontal zone, the relevant cross-frontal length scale is the half-width, $L/2$, yielding a geometric scaling of $\bar{u}_y \sim 2(H/L)m$. Similarly, for the available potential energy reservoir, the geometric isopycnal slope is defined as $S = -\bar{b}_y / Z$. Because the shallow aspect ratio of the domain ($(H/L)^2 \ll 1$) strictly dictates that the constant background stratification overwhelmingly dominates the dynamic vertical anomaly ($Z = N^2 + \bar{b}_z \approx N^2$), we can rigorously map the horizontal gradient term $-\bar{b}_y/N^2$ directly to the finite-amplitude isopycnal slope, $S$. 

Having transformed the mean state gradients into the global parameters $m$ and $S$, we formalize the turbulent closure using algebraic inequalities. Recall that the total domain-averaged eddy energy is defined as $e = \langle \frac{1}{2} ( u'^2 + v'^2 + (b'/N)^2 ) \rangle$. 

For the kinetic energy sink, the pointwise inequality $|u'v'| \le \frac{1}{2}(u'^2 + v'^2)$ ensures its domain average is bounded by the total eddy energy, such that $|\langle u'v' \rangle| \le e$. Given our assumed stable barotropic configuration, the turbulent momentum flux structurally phase-locks with the mean shear, continuously performing work to intensify the primary jet. Combining this energetic bound with our geometric scaling for the lateral shear, we parameterize this global domain-integrated extraction directly as $-\beta \frac{H}{L} m e$. Here, the non-dimensional barotropic phase efficiency $\beta \in [0, 1]$ naturally absorbs the $\mathcal{O}(1)$ geometric factor arising from the half-width length scale.

Next, for the lateral APE extraction, the analogous pointwise inequality $|v'(b'/N)| \le \frac{1}{2}(v'^2 + (b'/N)^2)$ guarantees its domain average is also bounded by the total eddy energy, such that $|\langle v'b' \rangle| / N \le e$. Because the eddies systematically act to flatten the isopycnals, the turbulent buoyancy flux structurally phase-locks to the orientation of the mean density gradient. We can therefore rigorously parameterize this flux using a non-dimensional structural efficiency $\alpha \in [0, 1]$. Applying this phase-locked bound to our isopycnal slope parameterization yields the lateral APE extraction as exactly $\alpha N |S| e$. This preserves dimensional consistency and guarantees a strictly positive-definite energy source, regardless of the front's dynamic orientation.

Combining these symmetric parameterizations yields the fully closed, dimensional ordinary differential equation for the total eddy energy:
\begin{equation}\label{eq:e_dot_closed_dimensional}
    \frac{de}{dt} = \left(-\beta \frac{H}{L}m + \alpha N |S| \right)e \, .
\end{equation}

\section{Derivation of the Domain-Averaged ODEs and Spatial Symmetries}
\label{sec:derivation}

\subsection{Volume Integration and the Dimensional ODEs}

Before evaluating the spatial derivatives for the mean state equations, we apply dynamic approximations to the turbulent field consistent with the energetic truncation established in Section \ref{sec:eddy_energy}. As justified previously, whether driven by an upscale kinetic energy cascade at the submesoscale or far-field Rossby wave radiation in the quasi-geostrophic limit, eddy momentum convergence selectively accelerates the primary along-front jet. Because there is no equivalent mechanism to drive cross-frontal momentum convergence, we neglect the meridional eddy momentum fluxes acting on the overturning cell ($\langle v'v' \rangle \to 0$, $\langle v'w' \rangle \to 0$). Similarly, to align with our isolated lateral kinetic energy closure, we filter the vertical component of the Reynolds stress ($\langle u'w' \rangle \to 0$) from the mean momentum budget. Conversely, both the meridional ($\langle v'b' \rangle$) and vertical ($\langle w'b' \rangle$) eddy buoyancy fluxes are retained.

Applying these turbulent truncations directly to Eqs.~\eqref{eq:mean_tendencies} yields the simplified mean state equations:
\begin{subequations}\label{eq:mean_truncated}
\begin{align}
    \frac{\partial \bar{u}}{\partial t} &= -\bar{\nabla} \cdot (\bar{\vec{v}}\bar{u}) + f\bar{v} - \frac{\partial}{\partial y}(\overline{u'v'}) \, , \label{eq:trunc_u} \\
    \frac{\partial \bar{v}}{\partial t} &= -\bar{\nabla} \cdot (\bar{\vec{v}}\bar{v}) - f\bar{u} - \frac{\partial \bar{\phi}}{\partial y} \, , \label{eq:trunc_v} \\
    \frac{\partial \bar{b}}{\partial t} &= -\bar{\nabla} \cdot (\bar{\vec{v}}\bar{b}) - N^2 \bar{w} - \frac{\partial}{\partial y}(\overline{v'b'}) - \frac{\partial}{\partial z}(\overline{w'b'}) \, . \label{eq:trunc_b}
\end{align}
\end{subequations}

Toward deriving the domain-averaged evolution, we apply local spatial derivatives ($\partial_i$, representing either $\partial_y$ or $\partial_z$) to the simplified mean PDEs (Eq.~\ref{eq:mean_truncated}), followed by the volume average $\langle \bar \cdot \rangle$ over the closed frontal domain. Because the nonlinear mean advection terms are already expressed in conservative flux form (utilizing mass continuity, $\bar{\nabla} \cdot \bar{\vec{v}} = 0$), the spatial derivative of any mean scalar $\xi$ expands as:
\begin{equation}
    \partial_i [ \bar{\nabla} \cdot (\bar{\vec{v}} \xi) ] = \bar{\nabla} \cdot (\bar{\vec{v}} \partial_i \xi) + (\partial_i \bar{\vec{v}}) \cdot \bar{\nabla} \xi \, .
\end{equation}
By Gauss's theorem, subject to the closed-domain boundary conditions, the volume integral of the first term on the right-hand side vanishes: $\langle \bar{\nabla} \cdot (\bar{\vec{v}} \partial_i \xi) \rangle = 0$. 

Consequently, we can bypass the full local continuous PDEs. The bulk evolution is governed purely by the surviving volume-averaged kinematic deformation ($\langle (\partial_i \bar{\vec{v}}) \cdot \bar{\nabla} \xi \rangle$), the Coriolis forces, and the double-derivative eddy flux divergences. To construct the equations for our domain-averaged variables, we systematically apply the vertical derivative ($-\partial_z$) to the cross-frontal momentum equation \eqref{eq:trunc_v}, the vertical derivative ($\partial_z$) to the along-front momentum equation \eqref{eq:trunc_u}, the lateral derivative ($-\partial_y$) to the buoyancy equation \eqref{eq:trunc_b}, and the vertical derivative ($\partial_z$) to the buoyancy equation \eqref{eq:trunc_b}. Evaluating these terms and explicitly expanding the dot products of the kinematic deformation yields the intermediate system:
\begin{subequations}\label{eq:macro_expanded}
\begin{align}
    -\frac{d}{dt}\langle \bar{v}_z \rangle &= \langle \bar{v}_z \bar{v}_y + \bar{w}_z \bar{v}_z \rangle + f\langle \bar{u}_z \rangle + \left\langle \frac{\partial^2 \bar{\phi}}{\partial y \partial z} \right\rangle \, , \label{eq:macro_exp_l} \\
    \frac{d}{dt}\langle \bar{u}_z \rangle &= -\langle \bar{v}_z \bar{u}_y + \bar{w}_z \bar{u}_z \rangle + f\langle \bar{v}_z \rangle - \left\langle \frac{\partial^2}{\partial y \partial z}(\overline{u'v'}) \right\rangle \, , \label{eq:macro_exp_m} \\
    -\frac{d}{dt}\langle \bar{b}_y \rangle &= \langle \bar{v}_y \bar{b}_y + \bar{w}_y \bar{b}_z \rangle + N^2 \langle \bar{w}_y \rangle + \left\langle \frac{\partial^2}{\partial y^2}(\overline{v'b'}) \right\rangle + \left\langle \frac{\partial^2}{\partial y \partial z}(\overline{w'b'}) \right\rangle \, , \label{eq:macro_exp_y} \\
    \frac{d}{dt}\langle \bar{b}_z \rangle &= -\langle \bar{v}_z \bar{b}_y + \bar{w}_z \bar{b}_z \rangle - N^2 \langle \bar{w}_z \rangle - \left\langle \frac{\partial^2}{\partial y \partial z}(\overline{v'b'}) \right\rangle - \left\langle \frac{\partial^2}{\partial z^2}(\overline{w'b'}) \right\rangle \, . \label{eq:macro_exp_z}
\end{align}
\end{subequations}

We now analyze the geometric limits and physical symmetries of these explicit equations to map them directly to our domain-averaged variables ($l, m, Y, Z$):

First, for the cross-frontal shear equation (Eq.~\ref{eq:macro_exp_l}), the exact mathematical cancellation of the mean nonlinear advection is revealed. Expanding the kinematic deformation yields $\langle \bar{v}_z \bar{v}_y + \bar{w}_z \bar{v}_z \rangle$. Substituting the 2D incompressibility condition ($\bar{w}_z = -\bar{v}_y$), this term evaluates identically to zero point-by-point. Next, by utilizing the hydrostatic balance ($\partial_z \bar{\phi} = \bar{b}$), the pressure cross-derivative is converted directly into the lateral buoyancy gradient: $\langle \partial_y(\partial_z \bar{\phi}) \rangle = \langle \bar{b}_y \rangle = -Y$. Finally, to formally map the surviving terms to our overturning vorticity variable $l = \langle \partial_y \bar{w} - \partial_z \bar{v} \rangle$, we evaluate the horizontal shear of the vertical velocity. Substituting mass continuity into the geometric scaling of the front dictates that $|\bar{w}_y| \sim \mathcal{O}(H^2/L^2)|\bar{v}_z|$. Because the aspect ratio is shallow, this term is geometrically suppressed, mathematically confirming that the vorticity is overwhelmingly defined by the meridional shear ($l \approx -\langle \bar{v}_z \rangle$). Substituting these elements, alongside the primary shear $m = \langle \bar{u}_z \rangle$, directly recovers the thermal wind imbalance for the closed domain-averaged vorticity:
\begin{equation}\label{eq:l_dot_dim}
    \frac{dl}{dt} = fm - Y \, .
\end{equation}

Second, for the primary along-front shear (Eq.~\ref{eq:macro_exp_m}): The explicitly arranged terms on the right-hand side carry direct physical interpretations. The expression $\langle \bar{v}_z (f - \bar{u}_y) \rangle$ represents the tilting of the absolute vertical vorticity into the circulation in the meridional plane. The second kinematic term, $-\langle \bar{w}_z \bar{u}_z \rangle$, represents the stretching of the vorticity in the meridional plane. Rather than applying an arbitrary dynamic filter to the submesoscale $\mathcal{O}(1)$ Rossby number regime, we note that the nonlinear components of these terms vanish naturally due to the spatial symmetries of the low-order projection. Assuming the jet is centered within the frontal zone, its horizontal shear $\bar{u}_y$ is anti-symmetric, yielding a vanishing domain average ($\langle \bar{u}_y \rangle = 0$). By decoupling the spatial averages for the primary mode ($\langle \bar{v}_z \bar{u}_y \rangle \approx \langle \bar{v}_z \rangle \langle \bar{u}_y \rangle$), the relative vorticity tilting drops out. Similarly, rigid horizontal boundary conditions dictate that the vertical average of $\bar{w}_z$ is exactly zero. Decoupling the spatial averages ($\langle \bar{w}_z \bar{u}_z \rangle \approx \langle \bar{w}_z \rangle \langle \bar{u}_z \rangle$) therefore eliminates the vorticity stretching term as well. 

Because these spatial symmetries hold, this low-order limit remains rigorously valid even for $\mathcal{O}(1)$ Rossby numbers. The averaged Coriolis torque thus maps directly to the macroscopic rotation ($f\langle \bar{v}_z \rangle = -fl$). Finally, we evaluate the Reynolds stress divergence, explicitly written as $\left\langle \frac{\partial}{\partial z} \left[ -\frac{\partial}{\partial y}(\overline{u'v'}) \right] \right\rangle$. This term represents the vertical gradient of the lateral eddy momentum convergence. Whether driven by far-field Rossby wave radiation (where wave activity converges aloft) or surface-trapped MLI (where submesoscale vortex merging maximizes near the boundary), the upscale transfer of kinetic energy dictates that lateral momentum convergence systematically increases with height. Consequently, this vertical gradient acts to intensify the mean vertical shear. Applying our energetic closure parameterization ($\langle |\overline{u'v'}| \rangle = \beta e$) and projecting this double-derivative operator onto the frontal geometry yields a positive kinematic forcing term, $\beta e / (HL)$, leaving the closed dimensional tendency:
\begin{equation}\label{eq:m_dot_dim}
    \frac{dm}{dt} = -fl + \frac{\beta e}{H L} \, .
\end{equation}

Third, for the horizontal buoyancy gradient (Eq.~\ref{eq:macro_exp_y}): Grouping the kinematic deformation yields $\langle \bar{v}_y \bar{b}_y \rangle + \langle \bar{w}_y (N^2 + \bar{b}_z) \rangle$. Substituting $Y = -\langle \bar{b}_y \rangle$ and $Z = N^2 + \langle \bar{b}_z \rangle$, and decoupling the spatial averages, the lateral divergence $\langle \bar{v}_y \rangle$ vanishes identically due to the no-normal flow boundary conditions at $y = \pm L/2$. The surviving kinematic term projects onto the macroscopic rotation: due to the geometric constraints of the overturning cell, the horizontal gradient of the vertical velocity $\langle \bar{w}_y \rangle$ is proportional to the macroscopic vorticity $l$, scaled by the aspect ratio compared to the vertical shear $\langle \partial_z \bar{v} \rangle$. This geometrically constrained coupling yields $\langle \bar{w}_y \rangle \approx (H/L)^2 l$, mapping directly to the kinematic production term $(H/L)^2 Zl$. Regarding the eddy flux divergences, applying the center-maximizing assumption and substituting our baroclinic closure ($\langle \overline{v'b'} \rangle = \alpha N e$) simplifies the lateral double-derivative to $-\alpha N e / L^2$. The mixed-derivative eddy term identically vanishes because $\overline{w'b'}$ is symmetric with respect to $y$, rendering its derivative anti-symmetric. Incorporating these simplifications leaves the closed dimensional tendency:
\begin{equation}\label{eq:y_dot_dim}
    \frac{dY}{dt} = \frac{H^2}{L^2}Zl - \frac{\alpha N e}{L^2} \, .
\end{equation}

Fourth, for the vertical stratification (Eq.~\ref{eq:macro_exp_z}): Grouping the background stratification and the dynamic anomaly yields the kinematic deformation $-\langle \bar{v}_z \bar{b}_y \rangle - \langle \bar{w}_z (N^2 + \bar{b}_z) \rangle$. Substituting $Z = N^2 + \langle \bar{b}_z \rangle$ and decoupling the spatial averages, the latter term depends on the domain-averaged vertical divergence, $\langle \bar{w}_z \rangle$. Whether bounded by rigid horizontal surfaces (as in a lake or ocean basin) or subject to radiation boundary conditions aloft (as in the atmosphere), this domain-averaged divergence vanishes identically ($\langle \bar{w}_z \rangle = 0$), completely eliminating the $-\langle \bar{w}_z \rangle Z$ term. The surviving kinematic deformation, $-\langle \bar{v}_z \bar{b}_y \rangle$, maps directly to the overturning vorticity without any geometric penalty. Since $l \approx -\langle \bar{v}_z \rangle$ and $Y = -\langle \bar{b}_y \rangle$, this evaluates exactly to $-Yl$. 

Regarding the eddy flux divergences, the mixed-derivative term $\langle \partial^2_{yz}(\overline{v'b'}) \rangle$ integrates to zero over the domain due to the meridional symmetry of the lateral buoyancy flux. The final vertical restratification term, $-\langle \partial^2_{zz}(\overline{w'b'}) \rangle$, scales geometrically as $+\langle \overline{w'b'} \rangle / H^2$. Constraining the eddy trajectories within the baroclinic wedge of instability maps the vertical buoyancy flux directly to the lateral flux. Because the eddies must release available potential energy and restratify the fluid regardless of whether the dense fluid resides to the north or south of the front, this conversion is proportional to the absolute isopycnal slope: $\langle \overline{w'b'} \rangle \approx (H/L) |S| \langle \overline{v'b'} \rangle = (H/L) |S| \alpha N e$. Substituting this yields a positive-definite kinematic forcing term, $\alpha N |S| e / (H L)$. Incorporating these simplifications leaves the closed dimensional tendency:
\begin{equation}\label{eq:z_dot_dim}
    \frac{dZ}{dt} = -Yl + \frac{\alpha N |S| e}{H L} \, .
\end{equation}

\subsection{Non-Dimensionalization and the Closed 5D System}

We non-dimensionalize this system using characteristic scales for length ($L$), depth ($H$), and horizontal velocity ($U$), denoting all non-dimensional variables with a tilde ($\tilde{\cdot}$). Time is scaled advectively, $\tilde{t} = t / (L/U)$, meaning the time derivative transforms as $d/dt = (U/L) d/d\tilde{t}$. The domain-averaged variables are normalized as: $\tilde{e} = e/U^2$, $\tilde{m} = m / (U/H)$, $\tilde{Y} = Y / (f U / H)$, and $\tilde{Z} = Z / N^2$. Because the overturning circulation is an ageostrophic secondary response, its vorticity scales with the Rossby number, $Ro \equiv U/(fL)$, yielding $\tilde{l} = l / (Ro \cdot U/H)$. 

We assume the horizontal spatial scale corresponds to the Rossby deformation radius, $L = L_d \equiv NH/f$. This inherently enforces the geometric ratio $H/L = f/N$. The dimensional isopycnal slope $S = Y/Z$ relates to the non-dimensional slope $\tilde{S} \equiv \tilde{Y}/\tilde{Z}$ via the algebraic identity $S = \tilde{S} (fU/H)/N^2$. Applying the deformation radius constraint, this ratio simplifies perfectly to $S = Ro (H/L) \tilde{S}$.

Substituting these scaled variables into the dimensional tendencies systematically absorbs all geometric constants. For instance, the kinematic ratio within the buoyancy flux simplifies strictly to $(L/H) |S| = Ro |\tilde{S}|$. Furthermore, when normalizing the vertical stratification equation, dividing the macroscopic kinematic forcing by the massive background static stability ($N^2$) reveals that the temporal evolution of $\tilde{Z}$ is dynamically suppressed by the squared Rossby number ($Ro^2$). Applying these relations to all five equations  reduces the Boussinesq framework to the closed, non-dimensional 5D dynamical system:
\begin{subequations}\label{eq:final_odes}
\begin{align}
    \dot{\tilde{l}} &= \frac{1}{Ro^2}(\tilde{m} - \tilde{Y}) \, , \\
    \dot{\tilde{m}} &= -\tilde{l} + \beta \tilde{e} \, , \\
    \dot{\tilde{Y}} &= \tilde{Z}\tilde{l} - \alpha \tilde{e} \, , \\
    \dot{\tilde{Z}} &= Ro^2 (-\tilde{Y}\tilde{l} + \alpha |\tilde{S}| \tilde{e}) \, , \\
    \dot{\tilde{e}} &= (\alpha |\tilde{S}| - \beta \tilde{m}) \tilde{e} \, .
\end{align}
\end{subequations}

\section{Dynamical Properties and Physical Interpretation of the Low-Order Model}
\label{sec:properties_interpretation}

\subsection{Constants of Motion and Phase-Space Geometry}

The dynamical system \eqref{eq:final_odes} admits two independent constants of motion. The first represents the total non-dimensional energy of the system, $\tilde{E}$:
\begin{equation}\label{eq:energy_cons}
    \tilde{E} = \frac{1}{2}\left[\tilde{m}^2 +(Ro \tilde{l})^2\right] -\frac{\tilde{Z}}{Ro^2} +\tilde{e} \, . 
\end{equation}
In the absence of eddies, this is equivalent to equation (8) in MH26 (where their scaling corresponds to the assumption $Ro \sim 1$). The first two terms represent the kinetic energy of the mean flow, where the first corresponds to the primary along-front jet and the second to the cross-frontal overturning circulation. The third term represents the mean potential energy. The last term, $\tilde{e}$, is the total eddy energy, comprising both kinetic and available potential components. (Note that the alternative form of energy conservation, given by equation (12) in MH26, is not recovered in this system due to the zonal uniformity condition ($\partial \bar{w}/\partial x = 0$), which renders the circulation in the $(x,z)$ plane unclosed.)

The second constant of motion relates directly to the thermodynamic structure of the front, defined by the square of the domain-averaged cross-frontal density gradient:
\begin{equation}\label{eq:Grad_Square}
    \tilde{\mathcal{R}}^2 = \tilde{Y}^2 + \left(\frac{\tilde{Z}}{Ro}\right)^2\, . 
\end{equation}
As established in Subsection~\ref{subsec:Bulk}, the dimensional magnitude of the macroscopic density gradient is proportional to $\sqrt{Y^2 + Z^2}$. Equation \eqref{eq:Grad_Square} serves as the non-dimensional analogue of this structural invariant, revealing a fundamental physical property of the model: the adiabatic dynamics governing the front, comprising both the mean overturning circulation and the parameterized eddy fluxes, cannot alter the overall magnitude of the density gradient. Instead, these processes act purely as rotational operators in the meridional plane. They adiabatically tilt the isopycnals, exchanging lateral buoyancy gradients ($\tilde{Y}$) for vertical stratification ($\tilde{Z}$). Thus, while the cross-frontal isopycnal slope, $\tilde{S} \equiv \tilde{Y}/\tilde{Z}$, evolves dynamically as the front steepens or slumps, the magnitude of the scaled density gradient $\tilde{\mathcal{R}}^2$ remains conserved.

By combining these two constants of motion and completing the square for $\tilde{Z}$, we can map the phase space geometry of the mean flow:
\begin{equation}\label{eq:hypersphere}
    \tilde{m}^2 + (Ro \tilde{l})^2 + \tilde{Y}^2 + \left(\frac{\tilde{Z}-1}{Ro}\right)^2 = \tilde{\mathcal{R}}^2 + \frac{1}{Ro^2} + 2(\tilde{E} - \tilde{e}) \, .
\end{equation}
Equation~\eqref{eq:hypersphere} reveals that the mean flow variables are geometrically bounded to the surface of a four-dimensional hypersphere. In the conservative, eddy-free limit ($\tilde{e}=0$), the radius of this hypersphere is constant, ensuring the trajectory of the mean flow remains bounded on a compact manifold. Because the envelope of this hypersphere constitutes a three-dimensional surface, the continuous-time dynamics possess the minimum degrees of freedom required to potentially exhibit chaotic behavior, consistent with the constraints of the Poincaré-Bendixson theorem \cite{strogatz2018nonlinear}. However, chaos has not been observed in this system. This regularity is presumably maintained by the continuously adjusting, inherent physical feedback mechanisms that regulate the flow, which will be detailed in the next subsection. When eddies are active ($\tilde{e} > 0$), these feedbacks cause the hypersphere's radius to shrink as eddy energy grows, geometrically restricting the amount of energy the mean flow can surrender to the turbulent field.

Although the system conserves total energy and the scaled density gradient, the parameterized eddy dynamics prevent it from satisfying Liouville's theorem \cite{goldstein2002classical}, as the phase-space flow is not volume-preserving. This is demonstrated by evaluating the phase-space divergence:
\begin{align}\label{eq:divergence}
    \sum_{i} \frac{\partial \dot{\tilde{x}}_i}{\partial \tilde{x}_i} &= \frac{\partial \dot{\tilde{l}}}{\partial \tilde{l}} + \frac{\partial \dot{\tilde{m}}}{\partial \tilde{m}} + \frac{\partial \dot{\tilde{Y}}}{\partial \tilde{Y}} + \frac{\partial \dot{\tilde{Z}}}{\partial \tilde{Z}} + \frac{\partial \dot{\tilde{e}}}{\partial \tilde{e}} \nonumber \\
    &= \alpha |\tilde{S}| \left(1 - \frac{Ro^2 \tilde{e}}{\tilde{Z}}\right) - \beta \tilde{m} \, .
\end{align}
It is important to note that all three terms on the right-hand side of Eq.~\eqref{eq:divergence} arise strictly due to the presence of the eddies. While this dependence is obvious for the term explicitly containing the eddy energy $\tilde{e}$ (which originates from the eddy-driven restratification in $\partial \dot{\tilde{Z}}/\partial \tilde{Z}$), the first ($\alpha |\tilde{S}|$) and third ($-\beta \tilde{m}$) terms also result entirely from the turbulent field, emerging directly from the growth rate of the eddy energy itself ($\partial \dot{\tilde{e}} / \partial \tilde{e}$). Because this divergence is generally non-zero, the phase-space volume contracts or expands depending on the flow state. This reflects the non-Hamiltonian nature of the eddy parameterization, despite the domain-averaged conservation of $\tilde{E}$ and $\tilde{\mathcal{R}}^2$.

\subsection{Physical Interpretation of the Inherent Feedbacks}
\label{subsec:physical}

Figure~2 illustrates a schematic snapshot of the dynamics governed by the closed 5D system~\eqref{eq:final_odes}. In this specific configuration, $\tilde{Y} > 0$, indicating that denser fluid is located to the north of the front. By hydrostatic balance, the pressure gradient force (PGF) in the upper layers is directed northward. This PGF drives a direct thermal cross-frontal circulation with a westward-pointing zonal vorticity. This primary forcing is represented by the $-\tilde{Y}$ term in the $\dot{\tilde{l}}$ tendency (Eq.~\ref{eq:final_odes}a).

Opposing this pressure gradient, the vertical shear of the along-front jet, $\tilde{m}$, exerts a differential Coriolis torque that acts to drive the cross-frontal circulation in the opposite (positive $\tilde{l}$) direction. The condition $\tilde{m} = \tilde{Y}$ represents the thermal wind balance, wherein the meridional PGF and the Coriolis force are in exact equilibrium. The quantity $(\tilde{m} - \tilde{Y})$ therefore represents the residual thermal wind imbalance. Rearranging Eq.~\eqref{eq:final_odes}a to $Ro^2 \dot{\tilde{l}} = \tilde{m} - \tilde{Y}$ immediately clarifies the role of inertia. In the highly rotationally constrained quasi-geostrophic (QG) limit ($Ro \to 0$), the inertial tendency vanishes, enforcing diagnostic thermal wind balance ($\tilde{m} \approx \tilde{Y}$). However, in the submesoscale $Ro \sim 1$ regime, non-negligible inertia permits this residual imbalance to drive a fully prognostic, time-dependent evolution of the cross-frontal circulation.

Equation~\eqref{eq:final_odes}b governs the along-front primary shear, $\tilde{m}$. The $-\tilde{l}$ term represents the direct feedback of the cross-frontal overturning circulation on the zonal flow via the Coriolis force (vorticity tilting). The second term, $\beta \tilde{e}$, parameterizes the barotropic convergence of zonal momentum driven by the eddies. As baroclinic instability generates turbulence, the upscale transfer of kinetic energy intensifies the mean shear. In the quasi-geostrophic limit ($Ro \ll 1$), this typically occurs as the outward radiation of eddy wave activity induces an inward flux of zonal momentum \cite{Vallis2017}. However, in the submesoscale $\mathcal{O}(1)$ Rossby number regime, this intensification can result directly from an inverse kinetic energy cascade driven by the merging of localized vortices. The constant parameter $\beta \ge 0$ dictates the overall efficiency of this upscale momentum transfer.

Equation~\eqref{eq:final_odes}c describes the evolution of the horizontal buoyancy gradient, $\tilde{Y}$. When the overturning circulation is thermally indirect ($\tilde{l}>0$, characteristic of a Ferrel-like cell), the advective circulation acts to steepen the isopycnals, increasing $\tilde{Y}$. Conversely, the term $-\alpha \tilde{e}$ represents the lateral eddy buoyancy flux. During active baroclinic instability ($\alpha > 0$), down-gradient eddy buoyancy transport acts to make northern latitudes denser and southern latitudes lighter, acting to flatten the isopycnals to reduce the mean available potential energy.

When considering both the eddy momentum forcing ($\beta \tilde{e}$) and the eddy buoyancy forcing ($-\alpha \tilde{e}$), eddy activity systematically pushes the system out of thermal wind balance. Defining the non-dimensional thermal wind imbalance residue as $\tilde{i} \equiv \tilde{m} - \tilde{Y}$, its temporal tendency evaluates to:
\begin{equation}\label{eq:imbalance_tendency}
    \dot{\tilde{i}} = (\alpha + \beta)\tilde{e} - (\tilde{Z}+1)\tilde{l} \, .
\end{equation}
The combined turbulent forcing, $(\alpha + \beta)\tilde{e}$, acts continuously to generate imbalance, while the cross-frontal overturning circulation, $\tilde{l}$, responds to restore it. 
In strictly quasi-balanced models, this imbalance tendency is assumed to remain negligible ($\dot{\tilde{i}} \approx 0$), yielding the diagnostic relation:
\begin{equation}\label{eq:eliptic}
    (\tilde{Z}+1)\tilde{l} = (\alpha + \beta)\tilde{e} \, .
\end{equation}
Equation~\eqref{eq:eliptic} serves as the direct low-order analogue of the spatial elliptic equations (e.g., the Sawyer-Eliassen equation \cite{Sawyer1956}). Furthermore, viewed through the lens of the Transformed Eulerian Mean (TEM) framework \cite{andrews1976planetary}, the combined turbulent forcing term $(\alpha + \beta)\tilde{e}$ acts as the projection of the Eliassen-Palm (EP) flux divergence, encompassing both the lateral eddy momentum flux ($\beta$) and the vertical eddy buoyancy flux ($\alpha$). Because the EP flux divergence is equal to the meridional eddy potential vorticity (PV) flux Eq.~\eqref{eq:eliptic} explicitly dictates how the overturning cell $\tilde{l}$ must instantaneously adjust to balance the total eddy PV flux.

Equation~\eqref{eq:final_odes}d governs the dynamic stratification anomaly, $\tilde{Z}$. Recalling that the total dimensional stratification is $Z = N^2 + \langle \bar{b}_z \rangle$, its non-dimensional form is $\tilde{Z} = 1 + \langle \bar{b}_z \rangle / N^2$. Thus, the temporal derivative $\dot{\tilde{Z}}$ explicitly tracks the evolution of the mean vertical buoyancy anomaly, dynamically normalized by the background state. In classical large-scale balanced dynamics ($Ro \ll 1$), the massive background stratification $N^2$ strictly dominates, and variations in the anomaly are considered negligible ($\dot{\tilde{Z}} \to 0$). However, in the $Ro \sim 1$ regime, this evolving component $\langle \bar{b}_z \rangle$ contributes non-negligibly to the total stratification, allowing the system to evolve interactively. The $-Ro^2 \tilde{Y} \tilde{l}$ term represents modification by the mean cross-frontal circulation, which advects denser fluid over lighter fluid, decreasing the static stability. The $Ro^2 \alpha |\tilde{S}| \tilde{e}$ term represents the vertical eddy buoyancy flux, which transports lighter fluid upward and denser fluid downward. This term extracts available potential energy to fuel the eddies while rigorously restratifying the fluid column.

By combining the temporal derivatives of $\tilde{Y}$ and $\tilde{Z}$, we derive an explicit evolutionary equation for the isopycnal slope $\tilde{S} \equiv \tilde{Y}/\tilde{Z}$. Assuming the standard frontal orientation where denser fluid resides to the north ($\tilde{S} > 0$, such that $\tilde{S}|\tilde{S}| = \tilde{S}^2$), the quotient rule neatly factors to:
\begin{equation}\label{eq:slope}
    \dot{\tilde{S}} = \left( 1 + Ro^2 \tilde{S}^2 \right) \left[ \tilde{l} - \frac{\alpha \tilde{e}}{\tilde{Z}} \right] \, .
\end{equation}
This mathematically manifests the direct physical competition at the front: baroclinic eddies ($\alpha \tilde{e} / \tilde{Z} > 0$) persistently act to flatten the isopycnal slope, whereas the thermally indirect cross-frontal circulation ($\tilde{l} > 0$) acts to steepen it.

Finally, Eq.~\eqref{eq:final_odes}e governs the total eddy energy, $\tilde{e}$. The $(\alpha |\tilde{S}|)$ term represents baroclinic growth, extracting available potential energy proportional to the magnitude of the isopycnal tilt. The $(-\beta \tilde{m})$ term represents barotropic stabilization, where the eddies perform work against the primary shear, transferring kinetic energy back to the mean flow. 

\section{Discussion}
\label{sec:discussion}

While this low-order 5D model is inherently a simplification of the full continuum dynamics, its compact nature allows us to transparently capture the interlocking, continuous adjustment feedbacks that regulate frontal systems. To illustrate this, we can trace the following physical scenario (summarized chronologically in Fig.~\ref{fig:dynamical_system}). 

Consider an initial state where the front is in exact thermal wind equilibrium ($\tilde{m} = \tilde{Y}$), with denser fluid situated to the north ($\tilde{Y} > 0$), as shown in Fig.~\ref{fig:dynamical_system}a. Assume the initial isopycnal slope is sufficiently steep to trigger active baroclinic instability. According to the eddy energy budget (Eq.~\ref{eq:final_odes}e), the turbulent field ($\tilde{e}$) begins to grow by extracting available potential energy. However, this growth establishes an immediate self-limiting feedback loop: as the eddies grow, their lateral buoyancy fluxes act to flatten the isopycnals (Eq.~\ref{eq:slope}), systematically cutting off the very baroclinic fuel that drives their growth. Concurrently, the eddies undergo barotropic stabilization, decaying by depositing kinetic energy into the mean flow. This momentum flux intensifies the along-front jet $\tilde{m}$ (Eq.~\ref{eq:final_odes}b). Because this barotropic decay rate depends directly on the magnitude of the primary shear itself, the intensification of the jet limits the further transfer of energy, effectively choking the inverse cascade (Eq.~\ref{eq:final_odes}e).

This turbulent activity inevitably disrupts the initial equilibrium (Fig.~\ref{fig:dynamical_system}b). The flattening of the isopycnals reduces the meridional buoyancy gradient ($\tilde{Y}$ decreases via Eq.~\ref{eq:final_odes}c), while the momentum deposition intensifies the jet ($\tilde{m}$ increases). As explicitly defined in Eq.~\ref{eq:imbalance_tendency}, this combined eddy forcing rapidly generates a positive thermal wind imbalance ($\tilde{i} = \tilde{m} - \tilde{Y} > 0$). 

The macroscopic mean flow responds dynamically to this imbalance. The intensified jet ($\tilde{m}$) generates an excess Coriolis torque, while the weakened lateral density gradient ($\tilde{Y}$) reduces the opposing pressure gradient force, whether that manifests as a weakened northward PGF in the upper atmosphere, or a weakened southward PGF in the lower layers of an ocean with a rigid lid. Together, these forces drive a thermally indirect, Ferrel-like overturning circulation ($\tilde{l} > 0$, Eq.~\ref{eq:final_odes}a; Fig.~\ref{fig:dynamical_system}c). 

However, once triggered, this overturning circulation acts to restore the thermal wind balance. The cross-frontal flow $\tilde{l}$ exerts a Coriolis drag that decelerates the jet ($\tilde{m}$ decreases via Eq.~\ref{eq:final_odes}b). Simultaneously, the indirect circulation drives adiabatic upwelling in the denser northern sector and subsidence in the lighter southern sector. This kinematic overturning cools the north and warms the south, acting to re-steepen the isopycnals, recover the lateral buoyancy gradient $\tilde{Y}$ (Eq.~\ref{eq:final_odes}c), and pull the system back toward equilibrium (Fig.~\ref{fig:dynamical_system}d).

Finally, these dynamics are intrinsically coupled to the evolving vertical stratification, $\tilde{Z}$. The baroclinic eddies transport lighter fluid upward and denser fluid downward, acting to increase the stratification (the $\alpha |\tilde{S}| \tilde{e}$ term in Eq.~\ref{eq:final_odes}d). However, because the vertical eddy buoyancy flux scales with the squared Rossby number ($Ro^2$), this effect competes with the macroscopic overturning circulation. For an indirect cell ($\tilde{l} > 0$) acting on a standard front ($\tilde{Y} > 0$), the circulation advects denser northern fluid over lighter southern fluid, actively reducing the static stratification. This, in turn, creates a nonlinear thermodynamic brake on the system: as the stratification $\tilde{Z}$ decreases, the vertical upwelling and downwelling motions become less efficient at modifying the lateral buoyancy gradient $\tilde{Y}$ (via the $\tilde{Z}\tilde{l}$ term in Eq.~\ref{eq:final_odes}c). Thus, the adiabatic overturning dynamics inherently limit their own capacity to indefinitely steepen the front, resulting in a newly balanced state that is broader and weaker than the initial configuration (Fig.~\ref{fig:dynamical_system}d).

\section{Conclusion and Future Work}
\label{sec:conclusion}

In this study, we aimed to derive a closed 5D low-order dynamical system, capable of representing the fundamental adiabatic adjustment of a frontal zone. By systematically volume-integrating the continuous Boussinesq equations and applying a domain-averaged eddy closure, we  projected the infinite-dimensional continuum dynamics onto a compact ordinary differential equation framework. This idealized system maintains two rigorous constants of motion: the total energy and the domain-averaged cross-frontal density gradient. Despite being purely conservative, lacking explicit external forcing or dissipation, the parameterization of the turbulent eddies renders the system non-Hamiltonian. The phase-space volume continuously expands and contracts as the eddies extract resources from, and deposit momentum into, the mean flow. 

This mathematical model situates the present system within a class of conceptual low-order models designed to elucidate atmospheric and oceanic dynamics. Much like the classical Lorenz (1984) model of jet-eddy interactions \cite{lorenz1984irregularity}, or the Ambaum and Novak (2014) nonlinear oscillator that frames storm-track variability as an ecological predator-prey relationship between baroclinicity and eddy heat flux \cite{ambaum2014nonlinear}, our 5D model attempts to distill complex continuous turbulence into its essential domain-integrated feedback loops.

Looking forward, this conservative framework may serve as the analytical description for several immediate extensions. First, while the discussion highlighted the system's robust inherent tendency to restore thermal wind balance, this restoring mechanism is not absolute. In a follow-up study, we will demonstrate that even strictly within the quasi-geostrophic (QG) limit, this system can bifurcate, transitioning abruptly between QG thermal wind adjustment and spontaneous, runaway imbalance depending on the initial conditions. Second, we will extend this framework into the submesoscale $Ro \sim 1$ limit to theoretically capture the fundamental dynamics of Mixed Layer Instability (MLI) \cite{boccaletti2007mixed}.

Finally, we aim to transition the model from a purely conservative to a forced-dissipative regime. Generally, the thermal wind balance is continuously perturbed by a myriad of external forcings and non-conservative processes. Atmospheric and oceanic frontogenesis is frequently driven by large-scale deformation and strain fields that act to continuously concentrate lateral buoyancy gradients \cite{hoskins1972atmospheric, hoskins1982mathematical}. Furthermore, these established fronts are heavily modulated by boundary interactions; they are subjected to diabatic heating, surface buoyancy fluxes, wind-driven Ekman transport \cite{thomas2005destruction}, and intense boundary-layer turbulence, such as wave-driven Langmuir circulations and convective mixing \cite{mcwilliams1997langmuir, belcher2012global}. Incorporating these mechanical and thermodynamic forcings into the 5D ODE framework will allow us to assess how atmospheric and oceanic baroclinic fronts reach statistical equilibrium in more realistic scenarios.

\begin{acknowledgments}
E.H. is grateful to Leo R. M. Maas for fruitful and enlightening discussions.
\end{acknowledgments}

\bibliography{references}

\begin{figure}[htbp]
    \centering
    \includegraphics[width=0.85\textwidth]{Slope_Struc.png}
    \caption{Schematic of the cross-frontal density structure in the meridional-vertical $(y, z)$ plane. The domain is bounded by $y \in [-L/2, L/2]$ and $z \in [0, H]$. The dashed blue lines indicate constant density surfaces (isopycnals) ${\overline\rho}$ and ${\overline\rho} \pm \Delta{\overline\rho}$, where ${\overline\rho}$ is the density at the geometric center $(y,z)_{g.c.}$. The total density gradient $\nabla \overline{\rho}$ (red arrow) is shown perpendicular to the isopycnals, decomposed into its horizontal ($\tilde{Y} > 0$) and vertical ($\tilde{Z} > 0$) components. The center of mass $(y,z)_{c.m.}$ is shifted relative to the geometric center due to the linear density distribution. The isopycnal slope in this example is $S = \tilde{Y}/\tilde{Z} = 0.60$.}
    \label{fig:Slope_Struc}
\end{figure}

\begin{figure*}[t]
    \centering
    \begin{subfigure}[t]{0.495\textwidth}
        \centering
        \begin{overpic}[height=4.5cm]{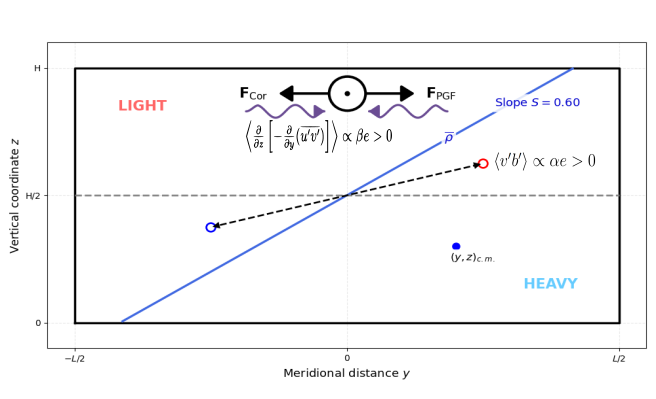}
            \put(2,60){\textbf{(a)}}
        \end{overpic}
        \label{fig:dyn_a}
    \end{subfigure}%
    \hspace{2pt}%
    \begin{subfigure}[t]{0.495\textwidth}
        \centering
        \begin{overpic}[height=4.5cm]{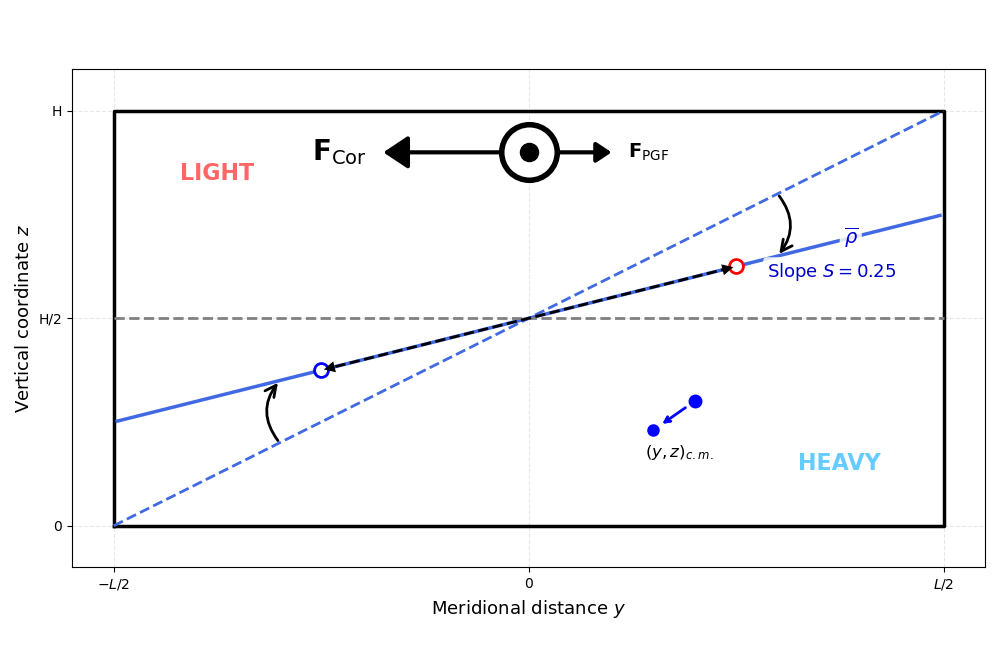}
            \put(2,60){\textbf{(b)}}
        \end{overpic}
        \label{fig:dyn_b}
    \end{subfigure}
    \\[-8pt]
    \begin{subfigure}[t]{0.495\textwidth}
        \centering
        \begin{overpic}[height=4.5cm]{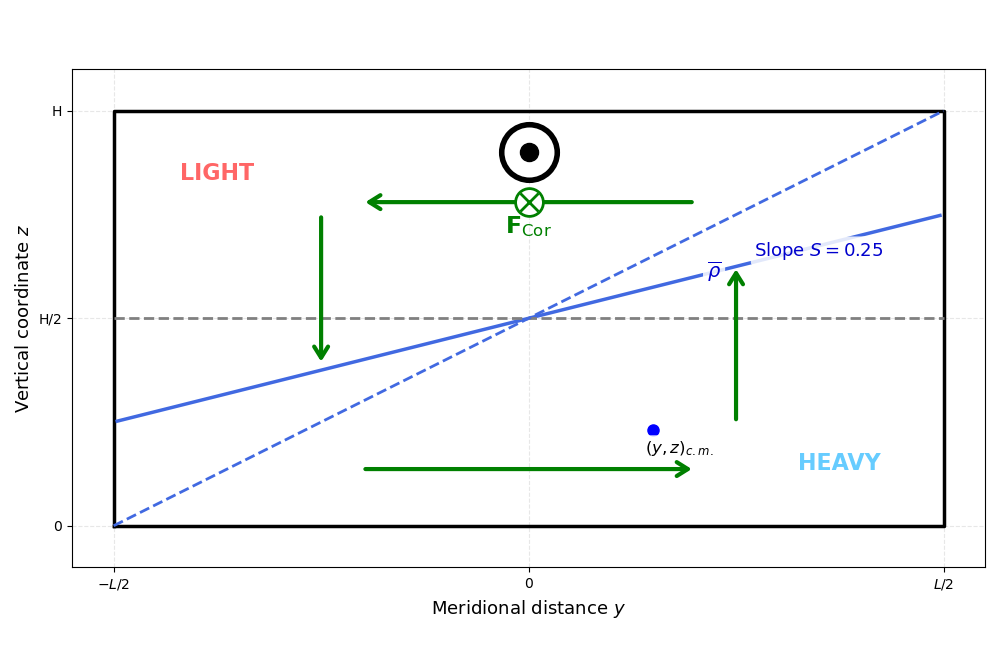}
            \put(2,60){\textbf{(c)}}
        \end{overpic}
        \label{fig:dyn_c}
    \end{subfigure}%
    \hspace{2pt}%
    \begin{subfigure}[t]{0.495\textwidth}
        \centering
        \begin{overpic}[height=4.5cm]{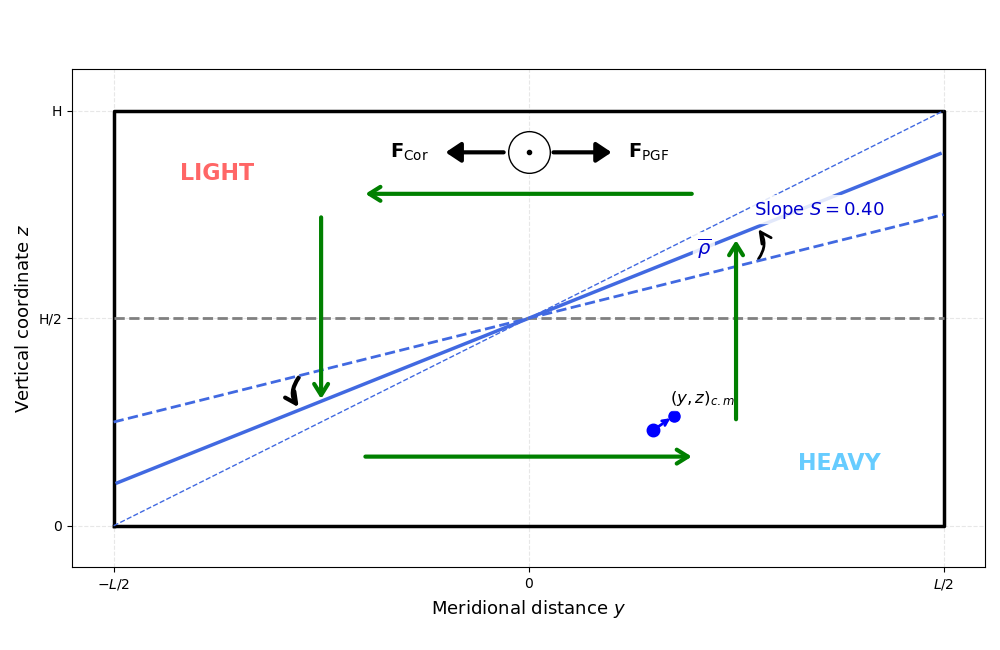}
            \put(2,60){\textbf{(d)}}
        \end{overpic}
        \label{fig:dyn_d}
    \end{subfigure}
    \caption{\textbf{Schematic of thermal wind adjustment and self-regulating feedbacks (Eq.~\ref{eq:final_odes}).} \textbf{(a) Initial Balanced State.} A meridional-vertical cross-section shows a surface-intensified frontal jet (cross/dot markers) in thermal wind equilibrium ($\tilde{m}=\tilde{Y}$). Pre-existing eddies ($\tilde{e}$) are embedded. \textbf{(b) Instability Growth and Imbalance.} Eddies consume available potential energy ($\dot{\tilde{e}} \sim \alpha |\tilde{S}| \tilde{e}$), flattening isopycnals to deplete their own resources. Concurrently, barotropic decay deposits momentum, intensifying the jet ($\tilde{m}$ increases). This generates a thermal wind imbalance ($\tilde{m} > \tilde{Y}$). \textbf{(c) Restorative OVC.} The imbalance triggers a thermally indirect overturning circulation ($\tilde{l}>0$, closed arrows). This restores balance via interlocking pairs: $\tilde{l}$ exerts a Coriolis drag that decelerates the jet ($\dot{\tilde{m}} \sim -\tilde{l}$, Eq.~\ref{eq:final_odes}b), and drives adiabatic vertical motion that re-steepens the isopycnals ($\dot{\tilde{Y}} \sim \tilde{Z}\tilde{l}$, Eq.~\ref{eq:final_odes}c). \textbf{(d) Re-Balancing.} Restoring feedbacks drive the system back toward equilibrium ($\tilde{m} - \tilde{Y} \to 0$), albeit to a weaker state than the initial balance. The jet broadens, while the OVC uploads dense fluid over light, reducing vertical stratification ($\dot{\tilde{Z}} < 0$, Eq.~\ref{eq:final_odes}d). This acts as a nonlinear thermodynamic brake on re-steepening.}
    \label{fig:dynamical_system}
\end{figure*}

\end{document}